\documentclass[preprint]{aastex62}

\usepackage{color}
\usepackage{amsmath}
\usepackage{bm}

\newcommand{\tred}[1]{#1}

\begin{document}

\title{
Effect of morphological asymmetry
between leading and following sunspots \\
on the prediction of solar cycle activity
}

\correspondingauthor{H. Iijima}
\email{h.iijima@isee.nagoya-u.ac.jp}

\author{H. Iijima}
\affiliation{
Center for Integrated Data Science,
Institute for Space-Earth Environmental Research, Nagoya University,
Furocho, Chikusa-ku, Nagoya, Aichi 464-8601, Japan.
}

\author{H. Hotta}
\affiliation{
Department of Physics, Faculty of Science, Chiba University,
1-33 Yayoi-chou, Inage-ku, Chiba 263-8522, Japan.
}

\author{S. Imada}
\affiliation{
Division for Integrated Studies,
Institute for Space-Earth Environmental Research, Nagoya University,
Furocho, Chikusa-ku, Nagoya, Aichi 464-8601, Japan.
}

\begin{abstract}
  The morphological asymmetry of leading and following sunspots is a well-known characteristic of the solar surface.
  In the context of large-scale evolution of the surface magnetic field, the asymmetry has been assumed to have only a negligible effect.
  Using the surface flux transport model, we show that the morphological
  asymmetry of leading and following sunspots has a significant impact on the evolution of the large-scale magnetic field on the solar surface.
  By evaluating the effect of the morphological asymmetry of each bipolar magnetic region (BMR), we observe that the introduction of the asymmetry in the BMR model significantly reduces its contribution to the polar magnetic field, especially for large and high-latitude BMRs.
  Strongly asymmetric BMRs can even reverse the regular polar field formation.
  The surface flux transport simulations based on the observed sunspot record shows that the introduction of the morphological asymmetry reduces the root-mean-square difference from the observed axial dipole strength by 30--40 percent.
  These results indicate that the morphological asymmetry of leading and following sunspots has a significant effect on the solar cycle prediction.
\end{abstract}



\section{Introduction}

The formation process of the polar magnetic field on the solar surface is important for understanding and predicting the long-term solar magnetic activity.
The polar magnetic field at cycle minimum reveals a strong correlation with the amplitude of the next solar cycle
\citep{1978GeoRL...5..411S,2005GeoRL..32.1104S,2005GeoRL..3221106S,2009ApJ...694L..11W,2013PhRvL.111d1106M}.
The precursor method based on the polar field observation is one of the few promising methods for solar cycle prediction \citep{1999JGR...10422375H,2010LRSP....7....6P,2012SoPh..281..507P}.
Prediction using the polar field has been extended with the surface flux transport model that predict the polar magnetic field based on the sunspot data \citep{2014ApJ...780....5U,2016ApJ...823L..22C,2016JGRA..12110744H,2017A&A...607L...2I,2018ApJ...863..159J}.
An accurate and long-term prediction of the polar magnetic field is desired for solar cycle prediction.

The morphological asymmetry between leading and following sunspots is a well-known feature observed in the solar photosphere \citep[][as cited in \cite{1961ApJ...133..572B}]{1950ZA.....28...28G}.
The morphological asymmetry has been observed in various aspects such as spatial size and shape \citep[e.g.,][]{1981NASSP.450..163Z}, sunspot area \citep[e.g.,][]{2014SoPh..289..563M,2014SoPh..289.1403T,2015AdSpR..55..835T}, magnetic field \citep[e.g.,][]{2017Ge&Ae..57..946Z}, spatial separation \citep[e.g.,][]{1990SoPh..126..285V,1998ASPC..140..105C}, and sometimes with the geometrical asymmetry of flux tube (e.g., east-west inclination) \citep{2009LRSP....6....4F}.
The physical origin of this asymmetry has been studied in the context of magnetic flux emergence and active region formation, using a thin flux tube model \citep[e.g.,][]{1993ApJ...405..390F,1995ApJ...441..886C,2011ApJ...741...11W} and magnetohydrodynamic simulation \citep[e.g.,][]{2008ApJ...676..680F,2014ApJ...785...90R,2017ApJ...846..149C}.
For the discussion and explanation of morphological asymmetry, please refer to \cite{2000SoPh..192..119F} and \cite{2009LRSP....6....4F}.

\begin{figure}[!ht]
\epsscale{0.8}
\plotone{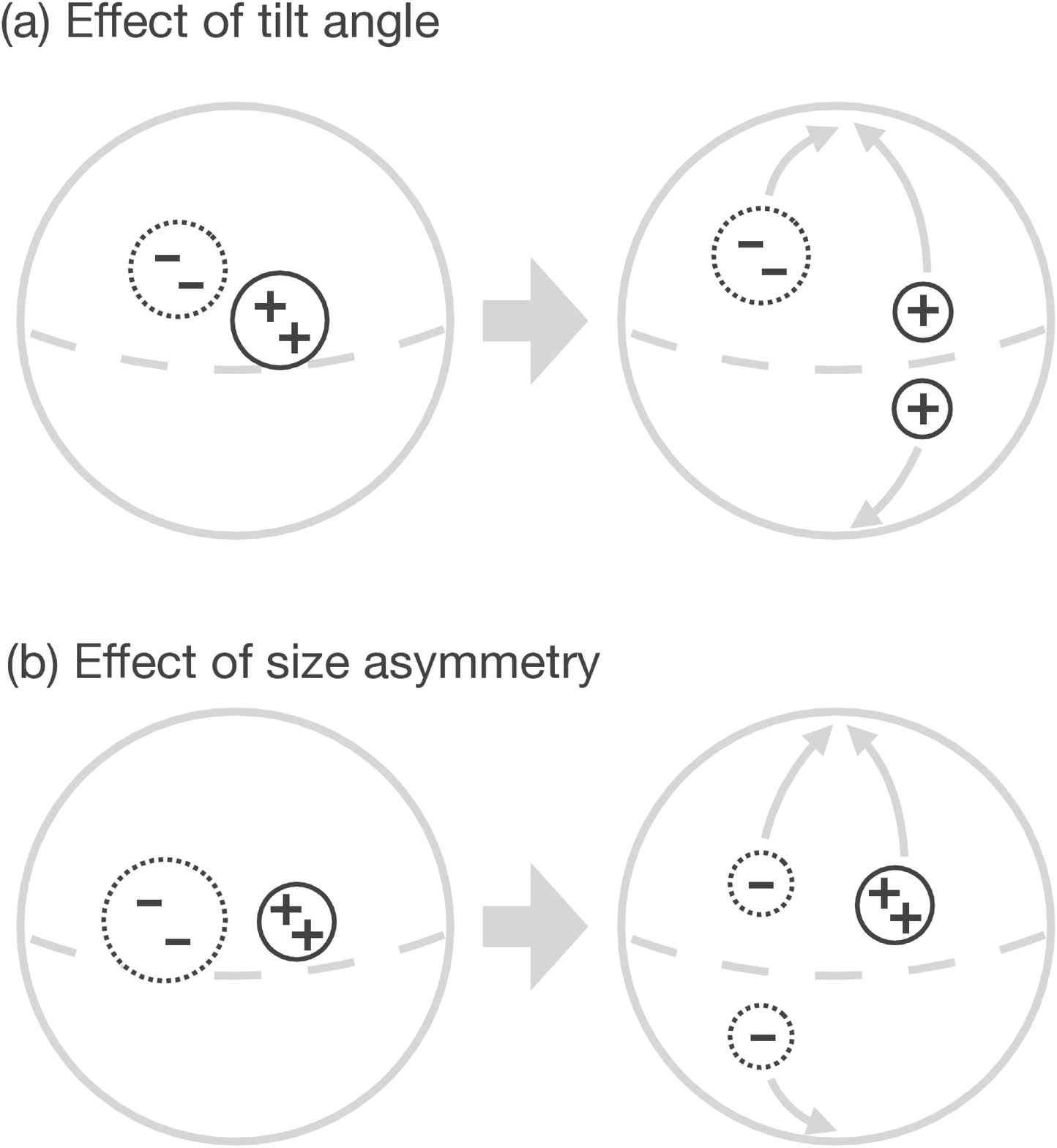}
\caption{
 Schematic illustration of the effects of the sunspot tilt angle (panel a) and size asymmetry (panel b) on the cross-equatorial flux transport and formation of the polar magnetic field.
 Thin gray arrows indicate the poleward transport by the meridional flow.
\label{fig:cartoon}
 }
\end{figure}

Morphological asymmetry was rarely studied in the context of solar cycle and polar field formation.
In this context, the latitudinal dependence of the tilt angle \citep[Joy's law;][]{1919ApJ....49..153H} has been studied as a key process of the polar field reversal.
In the standard view of the polar field formation (Fig. \ref{fig:cartoon}a), the tilt angle of the bipolar magnetic region (BMR) in the low latitude produces cross-equatorial transport in the magnetic flux of the leading polarity.
This cross-equatorial flux is advected and diffused to the polar region and contributes to the reversal and build-up of the polar magnetic field.

Meanwhile, we noted that the size asymmetry of sunspots pair in a BMR will also affect the polar field formation qualitatively (Fig. \ref{fig:cartoon}b).
The magnetic field distribution of the following sunspot is more diffuse (spatially wider) than that of the leading sunspot.
Thus, a spatially larger following magnetic patch is more likely to be transported across the equator than the leading patch.
This process will cancel the polarity of the cross-equatorial flux from the leading sunspots and will prevent the formation of the polar magnetic field.
As only a few studies have been reported on the effect of size asymmetry of sunspots on the formation of polar magnetic field, its quantitative effect is unknown.
In this study, we investigate the effect of size asymmetry of a sunspot pair in BMR on the formation of a polar magnetic field using the surface flux transport model.
We evaluate the dependence of the contribution to the polar magnetic field on the parameters of each BMR and quantify the impact on the surface field evolution based on the observed sunspot records.

The rest of the paper is organized as follows.
In Section \ref{sec:method} we describe our surface flux transport code and the model of BMR is introduced in the simulations.
We initiate the experiments on the asymmetry of a single BMR in Section \ref{sec:bmr}.
Section \ref{sec:conjec} is dedicated to conjecture the asymmetry parameter of the BMR model from the observed asymmetry of the leading and following sunspot areas.
Section \ref{sec:whole} describes the simulations of the multiple solar cycles and the effect of the sunspot area asymmetry with realistic parameters.
We summarize and discuss the results in Section \ref{sec:summary}.

\section{Surface flux transport model}
\label{sec:method}

We used the surface flux transport (SFT) model to approximate the evolution of the large-scale magnetic field on the solar photosphere.
In this study, a one-dimensional variant of the SFT model was used because the azimuthally averaged evolution of the surface magnetic field by the axisymmetric velocity and magnetic diffusion can be exactly reproduced by the one-dimensional SFT model \citep[e.g.,][]{2007ApJ...659..801C}.

\subsection{Basic equation}

The basic equation of the one-dimensional version of the SFT model can be written as
\begin{align}
 \frac{\partial B_R}{\partial t}
 +\frac{1}{R_\sun\sin\theta}
 \frac{\partial}{\partial\theta}
 \left(B_RV_\theta\sin\theta\right)
 =\frac{1}{R_\sun^2\sin\theta}
 \frac{\partial}{\partial\theta}
 \left(\eta_\mathrm{H}\sin\theta\frac{\partial B_R}{\partial\theta}\right)
 +S(\theta,t),
 \label{eq:basic}
\end{align}
where $\theta$ is the colatitude, $R_\sun$ is the radius of the sun ($\sim$ 696 Mm), $B_R$ is the radial component of the magnetic field, $V_\theta$ is the velocity  of the meridional flow, $\eta_\mathrm{H}$ is the coefficient of the horizontal turbulent diffusion, and $S(\theta,t)$ is the contribution of the emergence of new BMRs.
We used an analytical expression of the meridional flow described in \cite{1998ApJ...501..866V}.
The turbulent magnetic diffusivity was set to $250$ km$^2$/s \citep[e.g.,][]{2016ApJ...823L..22C}.
The above equations were solved using the second-order central difference method in space and the second-order strong stability preserving Runge--Kutta method in time \citep{gottlieb2009high}.
Equally spaced latitudinal grid of 0.176 deg (${\sim}2.14$ Mm) was used in all simulations described in this paper.

\subsection{Model of BMR}

The flux emergence term $S(\theta,t)$ is one of the greatest uncertainties of SFT models.
Various kinds of detailed structures of the flux emergence term have been suggested \citep{2014SSRv..186..491J}.
We follow the approach by \cite{2010ApJ...719..264C} and extend the model according to the size asymmetry of the leading and following sunspots.
In this paper, we describe the BMR model in a two-dimensional form.
The one-dimensional version of the BMR model can be derived by averaging each BMR in the longitudinal direction as described in Appendix \ref{appendix:1d_source}.

Each BMR is a sum of two magnetic patches with opposite polarities.
\begin{equation}
  B_\mathrm{BMR}(\theta,\phi)=
  s^{L}B^{L}(\theta,\phi)
  +s^{F}B^{F}(\theta,\phi),
\end{equation}
where the overscripts L and F denote leading and following polarities, respectively, and $s^{L}$($=-s^{F}$) represents the sign of the leading magnetic polarity.
Following \cite{1998ApJ...501..866V}, the spatial profile of each magnetic polarity $M(=L,F)$ can be approximated by
\begin{equation}
  B^{M}(\theta,\phi)=B_\mathrm{max}^{M}
  \exp\left[-\frac{2\left(1-\cos\beta^{M}\right)}
  {\left(\delta^{M}\right)^2}\right]
  \label{eq:each_patch}
\end{equation}
and
\begin{equation}
  \cos\beta^{M}=
  \cos\theta\cos\theta^{M}
  +\sin\theta\sin\theta^{M}
  \cos\left(\phi-\phi^{M}\right),
\end{equation}
where $B_\mathrm{max}^{M}$ is the maximum strength of the magnetic field, $\delta^{M}$ is the spatial size in radians, and $\beta^{M}$ is the distance from the center of the polarity.

The position of two magnetic patches in the BMR are defined as
\begin{equation}
  \theta^{L}=
  \theta_\mathrm{c}+{0.5}s_{\lambda}\varDelta\beta\sin\alpha,
\end{equation}
\begin{equation}
  \theta^{F}=
  \theta_\mathrm{c}-{0.5}s_{\lambda}\varDelta\beta\sin\alpha,
\end{equation}
\begin{equation}
  \phi^{L}=
  \phi_\mathrm{c}+{0.5}\varDelta\beta\cos\alpha(\sin\theta)^{-1},
\end{equation}
\begin{equation}
  \phi^{F}=
  \phi_\mathrm{c}-{0.5}\varDelta\beta\cos\alpha(\sin\theta)^{-1},
\end{equation}
where $s_{\lambda}$ is the sign of the latitude $\lambda=\pi/2-\theta$, $(\theta_\mathrm{c},\phi_\mathrm{c})$ is the central location of a BMR, $\alpha$ is the tilt angle with respect to the longitudinal direction, and $\varDelta\beta$ is the separation between the two polarities.

The magnetic flux of each polarity can be approximated by
\begin{equation}
  \Phi^{M}{\sim}\pi{R_\sun^2}
  \left(\delta^{M}\right)^2{B_\mathrm{max}^{M}}
\end{equation}
when the spatial sizes of the sunspots are sufficiently smaller than the solar radius (Appendix \ref{appendix:1d_source}).
Neglecting the cancellation between the two polarities (when the BMR is introduced into simulations), the total magnetic flux of the BMR is defined by
\begin{equation}
  \Phi^\mathrm{tot}=\Phi^{L}+\Phi^{F}.
\end{equation}
The magnetic flux of the BMR should be balanced between the two polarities as
\begin{equation}
  B_\mathrm{max}^{L}\left(\delta^{L}\right)^2
  {\sim}B_\mathrm{max}^{F}\left(\delta^{F}\right)^2
\end{equation}
to avoid the generation of a magnetic monopole.

In summary, each BMR can be characterized by eight parameters: the date of appearance, central position $(\theta_\mathrm{c},\phi_\mathrm{c})$, inclination angle $\alpha$, separation between bipoles$\varDelta\beta$, total unsigned flux $\Phi=\Phi^{L}+\Phi^{F}$, Gaussian width of each polarity ($\delta^{L}$, $\delta^{F}$) (or the maximum magnetic field strength of each polarity ($B_\mathrm{max}^{L}$, $B_\mathrm{max}^{F}$)).

\section{Effect of asymmetry in single BMR}
\label{sec:bmr}

\tred{
\subsection{Contribution of a isolated Gaussian magnetic patch to the polar magnetic field}
\label{sec:single_gaussian}
}

As a first step to analyze the polar field build-up by a single imposed BMR at the northern hemisphere, we investigated the characteristics of a Gaussian magnetic patch, which is a building block of a BMR.
One Gaussian magnetic patch is characterized by three parameters: maximum field strength $B_\mathrm{max}$, width of Gaussian magnetic patch$\delta$, and location of the Gaussian center $(\phi_\mathrm{c},\lambda_\mathrm{c})$.
As we focus on the evolution of the azimuthally averaged surface magnetic field, the central longitude $\phi_\mathrm{c}$ of the magnetic patch does not affect the result.
The contribution to the polar magnetic field can be quantified by the (southern) net hemispheric flux \citep[e.g.,][]{2013A&A...557A.141C}:
\begin{equation}
  \Phi_\mathrm{SH}
  =\int_\mathrm{SH}B_{R}dA
  =R_\sun^2\int_{-\pi/2}^{0}\int_0^{2\pi}
  B_{R}(\lambda,\phi){d\phi}{\cos{\lambda}d\lambda}
\end{equation}
Practically, this quantity is not affected by the hemispheric asymmetry because the net flux in the northern hemisphere must be canceled by the net flux in the southern hemisphere when the magnetic monopole is absent.
In this study, we violated the divergence-free condition between the hemispheres only in a numerical test with a single Gaussian magnetic patch described in this section.
The southern net hemispheric flux was chosen to evaluate the cross-equatorial transport of the magnetic flux from the northern hemisphere
because it becomes positive if the positive polarity in the northern hemisphere is transported into the southern hemisphere.
We run each simulation for five years after inserting a Gaussian magnetic patch, and we used the net hemispheric flux at the final snapshot as the measure of the contribution to the polar field formation.
The time derivative of the net hemispheric flux indicates the magnetic flux transported across the equator per unit time.
By evaluating the net hemispheric flux at the final snapshot (starting from zero net hemispheric flux), one can determine the total magnetic flux transported across the equator to the opposite hemisphere, which is approximately proportional to the contribution of the BMR on the axial dipole strength or the polar magnetic field strength several years after the BMR emergence.
The parameter survey on the maximum field strength $B_\mathrm{max}$ was avoided by normalizing the final net hemispheric flux with the magnetic flux of a single Gaussian magnetic patch (see Eq. \ref{eq:mn_flux}).

\begin{figure}[!ht]
\plotone{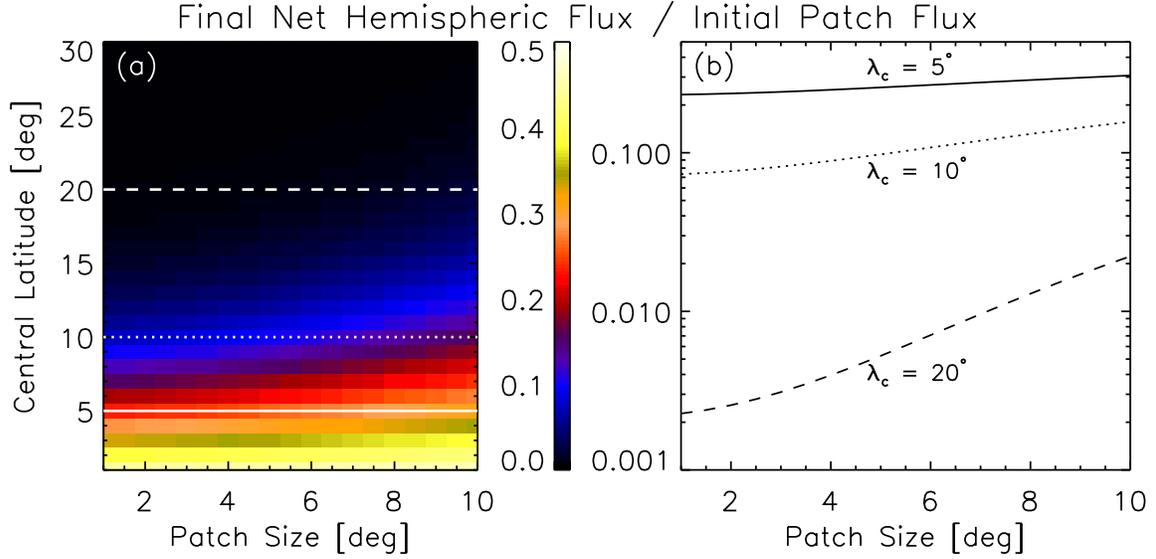}
\caption{
 Contribution of a Gaussian magnetic patch to the polar magnetic field (southern net hemispheric flux after five years that is normalized by the magnetic flux of initial patch).
 Panel a: normalized hemispheric flux as a function of the size $\delta$ and central latitude $\lambda_\mathrm{c}$ of the magnetic patch at the initial state.
 Horizontal white lines show the central latitude of 5 (solid), 10 (dotted), and 20 (dashed) [deg].
 Panel b: normalized hemispheric flux as a function of the initial patch size $\delta$ at the central latitude $\lambda_\mathrm{c}$ of 5 (solid), 10 (dotted), and 20 (dashed) [deg].
\label{fig:mn_mhf}
 }
\end{figure}
Figure \ref{fig:mn_mhf} presents the effect of the initial size $\delta$ of a single positive magnetic patch on the final net hemispheric flux of the five-year SFT run.
The result clearly indicates the importance of the initial size of a magnetic patch on the polar field formation.
The dependence on the initial patch size becomes stronger in the higher latitude; the ratio of the normalized hemispheric flux between $\delta=10$ [deg] and $\delta=1$ [deg] is 1.3, 2.1, and 9.8 at the central latitude of 5, 10, and 20 deg, respectively.
This result is contrary to the assumption used in several previous literatures\citep[e.g.,][]{1998ApJ...501..866V,2004A&A...426.1075B,2010ApJ...719..264C} that the initial size of the magnetic patch does not strongly affect the polar field build-up.

\tred{
\subsection{Effect of asymmetry in a pair of Gaussian magnetic patches (BMR)}
}

To characterize the size asymmetry \tred{between the leading and following polarities}, we used the square of the ratio of the size of the leading and following magnetic patches defined by
\begin{equation}
  f_\mathrm{\delta}=
  \left(\frac{\delta^{F}}{\delta^{L}}\right)^2
  {\sim}\ \frac{B_\mathrm{max}^{L}}{B_\mathrm{max}^{F}}.
\end{equation}
We assumed two relations to avoid the large number of free parameters to describe one BMR:
First, the tilt angle was defined by the central latitude of a BMR as $\alpha=g_\mathrm{inflow}T_n\sqrt{|\lambda|}$ following \cite{2010ApJ...719..264C}, where $g_\mathrm{inflow}=0.7$ is a factor to account for the active region inflow and $T_n=1.4$.
Second, the separation between the opposite polarities was set to be proportional to the spatial size of the leading polarity $\varDelta\beta=\delta^{L}/0.4=2.5\delta^{L}$, where the proportional constant 0.4 was chosen for maintaining consistency with previous studies \citep[e.g.,][]{1998ApJ...501..866V,2010ApJ...719..264C}.

\begin{figure}[!ht]
\epsscale{1.1}
\plotone{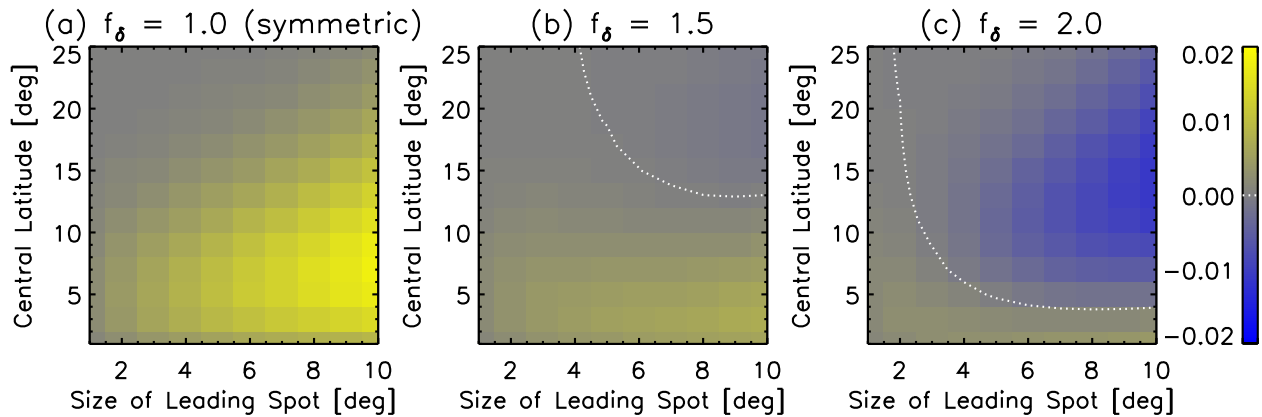}
\caption{
 Effect of the asymmetric sunspots on the contribution to the polar magnetic field by the emergence of a single BMR.
 Illustrated are the net hemispheric fluxes after five years normalization by the total (leading + following) unsigned magnetic flux of the BMR with (a) $f_\delta=1.0$ (symmetric case), (b) $f_\delta=1.5$ (weakly asymmetric case), and (c) $f_\delta=2.0$ (strongly asymmetric case).
 The white dotted lines indicate the location of zero contribution to the polar magnetic field.
\label{fig:bmr_mhf}
 }
\end{figure}
Figure \ref{fig:bmr_mhf} depicts the effect of the size asymmetry in a BMR on the formation of the polar magnetic field, measured by the net hemispheric flux after five years normalization by the total (leading + following) unsigned magnetic flux of the BMR.
We assumed positive polarity of the leading magnetic patch such that the positive (southern) net hemispheric flux indicates the net transport of leading polarity across the equator.
In the symmetric case ($f_\delta=1.0$; Fig. \ref{fig:bmr_mhf}a), a part of the magnetic flux of the leading polarity (several percents) is transported across equatorial region without being cancelled by the opposite polarity and contributes to the polar magnetic field.
The contribution becomes highest in the moderately low latitude of around 5--10 degree as the BMRs in the lower latitude have very small tilt angles.
\tred{
The size dependence of the cross-equatorial flux found in Sec. \ref{sec:single_gaussian} is also confirmed here for BMRs.
}
In the weakly asymmetric case ($f_\delta=1.5$; Fig. \ref{fig:bmr_mhf}b), the amount of the cross-equatorial flux becomes smaller than the asymmetric case, especially in higher latitudes ($\lambda_\mathrm{c}>15$ [deg]).
When the asymmetry becomes stronger ($f_\delta=2.0$; Fig. \ref{fig:bmr_mhf}c), the emergence of BMRs above $\lambda_\mathrm{c}>5$ [deg] prevents the natural formation of the polar magnetic field as the magnetic flux of the following polarity is transported across the equator.

The results also demonstrate that the spatially broader BMRs tend to be affected by the size asymmetry between leading and following polarities.
This is a natural consequence because the larger single magnetic polarity contributes towards the polar magnetic field with greater strength (Fig. \ref{fig:mn_mhf}).
We further carried out a simplified analysis on the diffusion process of a BMR (Appendix \ref{appendix:asymmetry}).
The analysis shows that the asymmetry of larger BMRs continues longer than those of smaller BMRs, which is also consistent with the strong dependence on the size of BMRs shown in Fig. \ref{fig:bmr_mhf}.

\section{Conjecture of the asymmetry parameter on the real Sun}
\label{sec:conjec}

In this section, we conjecture the possible value of the asymmetry parameter $f_\delta$ from the observational constraints, especially on the asymmetry of the sunspot area between the leading and following sunspots \citep{2014SoPh..289.1403T}.
To reach this goal, we define the ``sunspot area'' of a BMR approximated by two Gaussian magnetic patches.
Although the real sunspot does not exhibit such an idealized spatial shape, we use this simplified BMR model to avoid the large number of new free parameters not constrained by the observation.

\subsection{Definition of sunspot area in a single Gaussian magnetic patch}

\begin{figure}[!ht]
\plotone{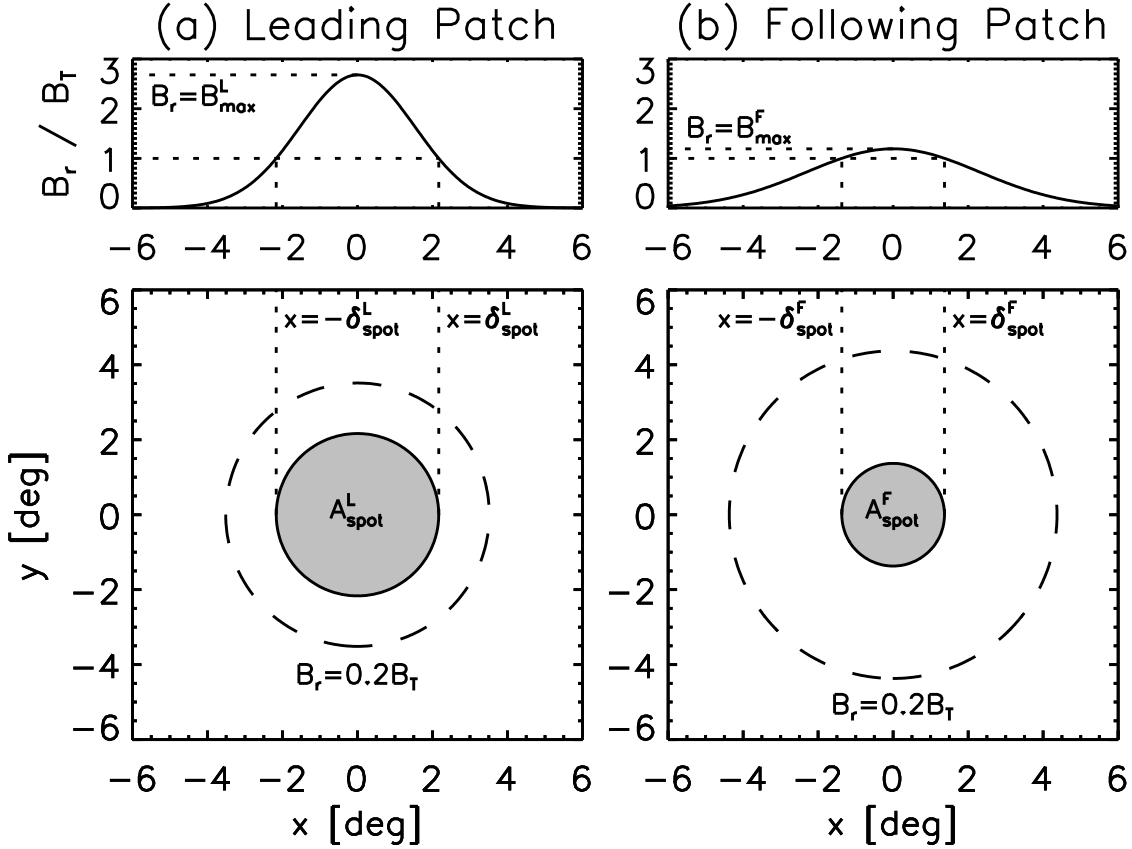}
\caption{
 Example of (a) leading and (b) following magnetic patch in the BMR model used in Sections \ref{sec:conjec} and \ref{sec:whole}.
 Top panels show the cross sections of the magnetic patches across each peak position.
 Bottom panels show the two-dimensional profile of the modeled magnetic patch.
 The region inside the sunspot ($B_r>B_\mathrm{T}$) is shaded with gray color.
 Dashed lines indicate $B_r=B_\mathrm{T}$.
 The parameters used in the BMR model are $f_\mathrm{spot}=0.4$, $c=2\mathrm{e}/(1+f_\mathrm{spot})$, and $A_\mathrm{spot}=1000$ [msh], which give the average sunspot field strength of $B_\mathrm{spot}=1.53B_\mathrm{T}$ and the parameter of the size asymmetry $f_\delta=2.25$.
\label{fig:bmr_2d}
 }
\end{figure}
First, we introduced the concept of ``sunspot area'' on a magnetic monopole region as a building block of a BMR (see Fig. \ref{fig:bmr_2d}).
We set a threshold strength of the magnetic field $B_\mathrm{T}$ and defined the ``sunspot'' of a Gaussian magnetic patch as a region where the magnetic field strength exceeds the threshold strength.
Given the spatial Gaussian width $\delta$ and the maximum field strength $B_\mathrm{max}$ in addition to the threshold value, we can derive the sunspot area $A_\mathrm{spot}$ and magnetic flux in the sunspot $\Phi$.
Inversely, if the total monopole magnetic flux (sunspot flux plus non-sunspot flux) and the sunspot area are given, we can compute the width and the peak field strength of the Gaussian patch.
For simplicity, we omitted the superscript $M=L,F$ in this subsection.

We used a functional form of a monopole magnetic patch (see Eq. \ref{eq:each_patch}) defined by
\begin{equation}
  B(r)=B_\mathrm{max}\exp(-r^2/\delta^2),
\end{equation}
where $r$ is the distance from the Gaussian center measured in unit of radian.
The total (sunspot plus non-sunspot) magnetic flux is given by
\begin{equation}
  \Phi=\pi{R_\sun^2}B_\mathrm{max}\delta^2.
\end{equation}
If we assume a round sunspot, the radius of the sunspot $\delta_\mathrm{spot}$ and the sunspot area should satisfy the relation
\begin{equation}
  A_\mathrm{spot}=\pi{R_\sun^2}\delta_\mathrm{spot}^2.
\end{equation}
The magnetic field strength at the edge of the sunspot $B(r=\delta_\mathrm{spot})$ should be equal to the threshold strength $B_\mathrm{T}$ as
\begin{equation}
  B_\mathrm{T}
  =B_\mathrm{max}\exp(-\delta_\mathrm{spot}^2/\delta^2).
\end{equation}

Eliminating the Gaussian width $\delta$ from the above relations, we can get an equation for the peak field strength $B_\mathrm{max}$ as
\begin{equation}
  B_\mathrm{max}
  \exp\left[-\frac{A_\mathrm{spot}B_\mathrm{max}}{\Phi}\right]
  =B_\mathrm{T}.
\end{equation}
The solution of the equation is given by
\begin{equation}
  B_\mathrm{max}
  =-\frac{\Phi}{A_\mathrm{spot}}
  W_0\left(-\frac{B_\mathrm{T}A_\mathrm{spot}}{\Phi}\right),
\end{equation}
where $W_0(x)$ is the principle branch of the Lambert W-function.
The spatial width of the Gaussian patch $\delta$ is given as
\begin{equation}
  \delta
  =\left(\frac{\Phi}{\pi{R_\sun^2}B_\mathrm{max}}\right)^{1/2}
  =\delta_\mathrm{spot}
  \left[-W_0\left(-\frac{B_\mathrm{T}
  A_\mathrm{spot}}{\Phi}\right)\right]^{-1/2}.
\end{equation}
If the total flux is proportional to the sunspot area as $\Phi/A_\mathrm{spot}=\mathrm{const.}$, $B_\mathrm{max}$ and $\delta/\delta_\mathrm{spot}$ do not depend on the actual value of the sunspot area $A_\mathrm{spot}$.
This characteristic is useful when we discuss the asymmetry of the sunspot area in the following sections.
From the definition range of the Lambert W-function, the relation
\begin{equation}
  0<\frac{B_\mathrm{T}A_\mathrm{spot}}{\Phi}
  {\le}\frac{1}{\mathrm{e}}
\end{equation}
should be satisfied to obtain the real positive value of $B_\mathrm{max}$ and $\delta$.
When this relation is satisfied, the allowed range of the solution is given by
\begin{equation}
  0<B_\mathrm{max}{\le}\Phi/A_\mathrm{spot}
  ,\
  \delta_\mathrm{spot}{\le}\delta
  .
\end{equation}
From the above procedure, we can obtain the width $\delta$ and peak value $B_\mathrm{max}$ of a Gaussian magnetic patch as a function of the sunspot area $A_\mathrm{spot}$, the total flux of the monopole$\Phi$, and the threshold magnetic field strength $B_\mathrm{T}$.

\subsection{Definition of sunspot area asymmetry}

We are now ready to discuss the relation between the asymmetry of the leading and following Gaussian magnetic patches in a BMR and the observationally constrained asymmetry of the leading and following sunspot area.
In the discussion below, we neglect the cancellation of opposite polarities for simplicity.
In this study, we defined the asymmetry parameter as a ratio between the areas of the leading and following sunspots
\begin{equation}
  f_\mathrm{spot}=
  \frac{A_\mathrm{spot}^{F}}{A_\mathrm{spot}^{L}}
  =\left(\frac{\delta_\mathrm{spot}^{F}}
  {\delta_\mathrm{spot}^{L}}\right)^2.
\end{equation}
The sum of the leading and following sunspot areas should be equal to the  sunspot group area as
\begin{equation}
  A_\mathrm{spot}=A_\mathrm{spot}^{L}
  +A_\mathrm{spot}^{F},
\end{equation}
which leads to
\begin{equation}
  A_\mathrm{spot}^{L}=\frac{A_\mathrm{spot}}{1+f_\mathrm{spot}}
  ,\
  A_\mathrm{spot}^{F}
  =\frac{f_\mathrm{spot}A_\mathrm{spot}}{1+f_\mathrm{spot}}.
\end{equation}

As an example, we discuss the case when the total (leading plus following) magnetic flux is proportional to the sunspot area as
\begin{equation}
  \Phi_\mathrm{tot}=cB_\mathrm{T}A_\mathrm{spot},
\end{equation}
where $c$ is a parameter that describes the amount of the total magnetic flux in a BMR relative to the lower bound of the magnetic flux in sunspots $B_\mathrm{T}A_\mathrm{spot}$.
The magnetic fluxes of leading and following polarities were assumed to be the equipartition $\Phi^{L}=\Phi^{F}=\Phi_\mathrm{tot}/2$.
From the definition range of the Lambert W-function, we can obtain the lower limit of the parameter $c$ as
\begin{equation}
  c{\hspace{0.1em}\ge\hspace{0.1em}}
  \frac{2\mathrm{e}}{1+f_\mathrm{spot}}
  {\equiv}c_\mathrm{min}
\end{equation}
assuming that the sunspot area of the leading polarity is equal to or greater than that of the following polarity as $f_\mathrm{spot}{\le}1$.

\begin{figure}[!ht]
\epsscale{0.7}
\plotone{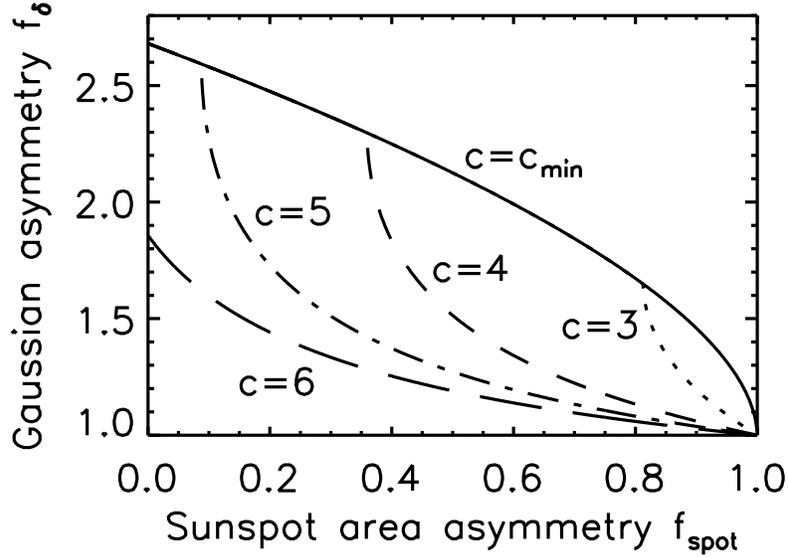}
\caption{
 Relation between the asymmetries of patch sizes in the BMR model and observed sunspot area.
 Each line represents a different value of $c=\Phi_\mathrm{tot}/(B_\mathrm{T}A_\mathrm{spot})$.
 Depicted are the cases with $c=c_\mathrm{min}=2\mathrm{e}/(1+f_\mathrm{spot})$ (solid), $c=3$ (dotted), $c=4$ (dashed), $c=5$ (dash-dotted), and $c=6$ (long-dashed).
\label{fig:fg_vs_fs}
 }
\end{figure}
Now we have acquired a tool to speculate the possible asymmetry of patch sizes in our BMR model from the asymmetry of the leading and following sunspot areas (Fig. \ref{fig:fg_vs_fs}).
As expected, the smaller following sunspot area produces wider Gaussian patch with the following polarity.
We note that the asymmetry of the patch size $f_\delta$ is a function of both the sunspot area asymmetry $f_\mathrm{spot}$ and the total flux factor $c$.
The smaller value of $c$ produces stronger asymmetry of $f_\delta$ for constant $f_\mathrm{spot}$.
With an observational value of the sunspot area asymmetry $f_\mathrm{spot}{\sim}0.4$ \citep{2014SoPh..289.1403T}, we obtain the BMR asymmetry $f_\delta{\sim}1.3-2.2$ for $c_\mathrm{min}{\le}c{\le}6$.
In the latter section, the possible value of $c$ will be determined by comparing the observed and simulated surface magnetic fields.

To use the observationally constrained value of the average magnetic field strength in a sunspot, we defined the average sunspot area in a BMR $B_\mathrm{spot}$.
The magnetic flux in the sunspot region of each polarity can be written as
\begin{equation}
  \Phi_\mathrm{spot}^{M}
  =
  \Phi^{M}\left[1-\mathrm{e}^{-\left(
  \delta_\mathrm{spot}^{M}/\delta^{M}
  \right)^2}\right]
  =
  \frac{\Phi_\mathrm{spot}}{2}
  \left(1-\frac{B_\mathrm{T}}{B_\mathrm{max}^{M}}\right)
\end{equation}
From the definition, $B_\mathrm{spot}$ can be derived as
\begin{equation}
  B_\mathrm{spot}
  =
  \frac{\Phi_\mathrm{spot}^{L}+\Phi_\mathrm{spot}^{F}}
  {A_\mathrm{spot}^{L}+A_\mathrm{spot}^{F}}
  =
  \frac{cB_\mathrm{T}}{2}
  \left(2-\frac{B_\mathrm{T}}{B_\mathrm{max}^{L}}
  -\frac{B_\mathrm{T}}{B_\mathrm{max}^{F}}\right).
\end{equation}
As $B_\mathrm{max}^{M}/B_\mathrm{T}$ is a function of $c$ and $f_\mathrm{spot}$, the ratio $B_\mathrm{spot}/B_\mathrm{T}$ depends only on $c$ and $f_\mathrm{spot}$.
\begin{figure}[!ht]
\plotone{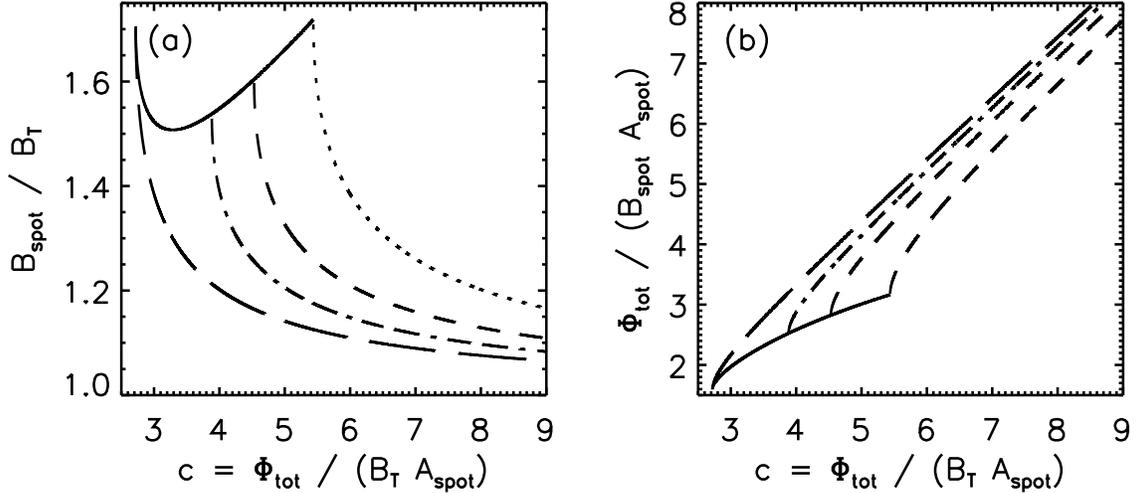}
\caption{
 Dependence of (a) the normalized average magnetic field strength in sunspots $B_\mathrm{T}/B_\mathrm{T}$ and (b) the ratio between the total magnetic flux and sunspot flux $\Phi_\mathrm{tot}/(B_\mathrm{spot}A_\mathrm{spot})$ on the total flux factor $c=\Phi_\mathrm{tot}/(B_\mathrm{T}A_\mathrm{tot})$.
 Depicted are $f_\mathrm{spot}=2\mathrm{e}/c-1$ (solid), $f_\mathrm{spot}=10^{-3}$ (dotted), $f_\mathrm{spot}=0.2$ (dashed), $f_\mathrm{spot}=0.4$ (dot-dashed), and $f_\mathrm{spot}=1.0$ (long-dashed).
\label{fig:bspot_bth}
 }
\end{figure}
Figure \ref{fig:bspot_bth} illustrates the parameter dependence of the normalized average strength of the sunspot magnetic field $B_\mathrm{spot}/B_\mathrm{T}$ (panel a) and the dependence of the ratio between the total magnetic flux and sunspot flux $\Phi_\mathrm{tot}/B_\mathrm{spot}A_\mathrm{spot}$ (panel b).
These quantities are used to constrain the free parameter $c$ from the observational data in the latter part of this paper.

\section{Effect of asymmetry in whole solar cycle}
\label{sec:whole}


To quantify the effect of the sunspot area asymmetry in the surface flux transport process, we carried out SFT simulations based on the observed sunspot record from Cycle 21 to 24.
We used the USAF/NOAA SOON data from \url{http://solarcyclescience.com/activeregions.html} to constrain our BMR model.
The sunspot group areas and locations were chosen at the timing of the maximum group area.
As shown by \cite{2015ApJ...800...48M}, the number of the small sunspot groups in the SOON data is erroneous when the group area is smaller than around 100 msh.
In the simulations of this study, we excluded the active region with the sunspot group area with less than 100 msh.
The tilt angle is determined following \cite{2010ApJ...719..264C} as in Section \ref{sec:bmr} except for using the cycle dependent factor $T_n$ to simulate multiple solar cycles \citep{2010ApJ...719..264C,2011A&A...528A..82J,2018ApJ...863..159J}.
The separation between polarities is modeled as in Section \ref{sec:bmr}.
The initial condition of the magnetic field is the quasi-steady solution proposed by \cite{1998ApJ...501..866V}.

\begin{table}[ht!]
\centering
\caption{
 Range and step size of free parameters used for the optimization of the multiple cycle simulations.
 }
\label{tab:grid_search}
\begin{tabular}{cccccc}
\hline
\hline
&$c-c_\mathrm{min}$
&$T_{21}$&$T_{22}$&$T_{23}$&$T_{24}$\\
\hline
\# of grids & 11 & 16 & 16 & 16 & 16 \\
  Max & 5.0 & 2.0 & 2.0 & 2.0 & 2.0 \\
  Min & 0.0 & 0.5 & 0.5 & 0.5 & 0.5 \\
  Step & 0.5 & 0.1 & 0.1 & 0.1 & 0.1 \\
\hline
\end{tabular}
\end{table}
Consequently, the free parameters that remained in our BMR model are the strength of the initial condition, threshold strength of the magnetic field $B_\mathrm{T}$, fraction of the non-spot magnetic flux $c$, and cycle dependent factor of the tilt angle $T_n$ for $n=$ 21, 22, 23, and 24.
Using the linearity of the SFT model on the magnetic field, we determined the strength of the initial condition and threshold strength of the magnetic field $B_\mathrm{T}$ by minimizing the L$_2$-norm of the difference between the simulated and observed axial dipole strengths.
The WSO synoptic magnetogram was used as the reference observation.
We optimized other free parameters using the grid search with the parameter range and step size given in Table \ref{tab:grid_search}.

We did not optimize the asymmetry parameter $f_\mathrm{spot}$.
Instead, we used $f_\mathrm{spot}=0.4$ as the typical value inferred from the observational data \citep{2014SoPh..289.1403T}.
The simulation with the symmetric spots ($f_\mathrm{spot}=1.0$) was also carried out for the purpose of comparison.

\begin{table*}[ht!]
\centering
\caption{
Optimal parameters and residual errors with $f_\mathrm{spot}=$ 0.4 and 1.0.
 }
\label{tab:optimal_solution}
\begin{tabular}{cccccccc}
\hline
\hline
  Cases &$B_\mathrm{T}$ [G]
&$c-c_\mathrm{min}$
&$T_{21}$&$T_{22}$&$T_{23}$&$T_{24}$
& Error\tablenotemark{a} [G] \\
\hline
$f_\mathrm{spot}=1.0$& 1830
& 0.0 & 1.0 & 1.4 & 1.1 & 1.9 & 3.50 \\
$f_\mathrm{spot}=0.4$& 1670
& 0.0 & 1.4 & 1.5 & 1.3 & 1.6 & 2.18 \\
\hline
\end{tabular}
\tablenotetext{a}{Root-mean-square error of the axial dipole strength from the WSO data.}
\end{table*}
The optimal parameters and residual errors are shown in Table \ref{tab:optimal_solution}.
We note that the resultant error of the axial dipole strength was several tens of percents smaller when the area asymmetry of sunspots is considered.
The threshold strength of the magnetic field, which indicates the magnetic field strength at the boundary of the sunspots, has an optimal value of approximately 1--2 kilo-gauss in both cases, which provides the physical consistency to our approach.

\begin{figure}[!ht]
\epsscale{0.8}
\plotone{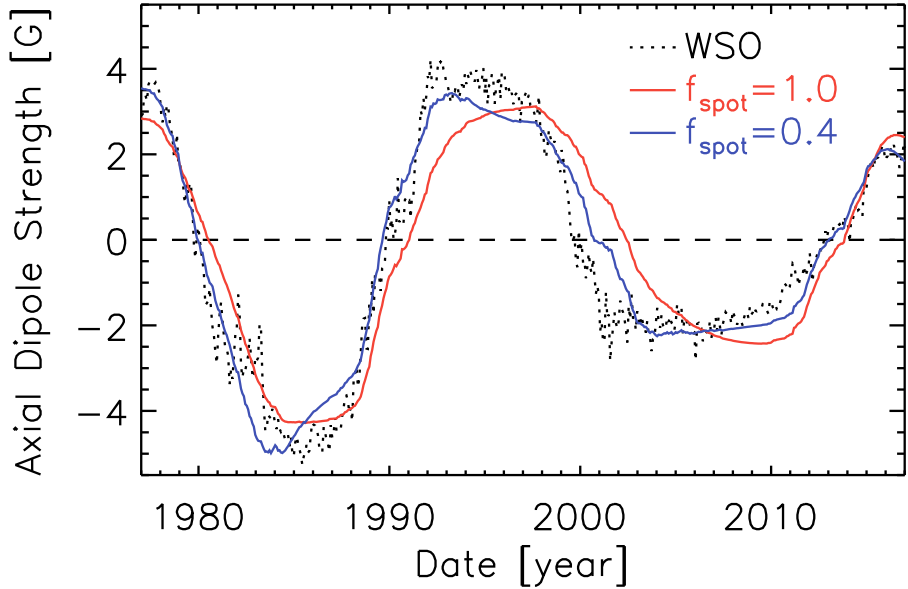}
\caption{
 Axial dipole strength of the optimal solutions shown in Table \ref{tab:optimal_solution}.
 Shown are the cases without asymmetry ($f_\mathrm{spot}=1.0$; red solid), with asymmetry ($f_\mathrm{spot}=0.4$; blue solid), and the observation by WSO (black dotted).
\label{fig:plmdp_cmp_rgo}
 }
\end{figure}
Figure \ref{fig:plmdp_cmp_rgo} shows the temporal evolution of the axial dipole strength in the optimal solutions.
With the effect of the area asymmetry ($f_\mathrm{spot}=0.4$), the timing of the axial dipole reversal is earlier than the symmetric case and closer to the observation.
Near the end of each cycle, the symmetric case ($f_\mathrm{spot}=1.0$) exhibited the continuous amplification of the axial dipole strength until the end of each cycle.
Meanwhile, the asymmetric case exhibited gradual decrease of the axial dipole strength several years before the end of each cycle.

\begin{figure}[!ht]
\plotone{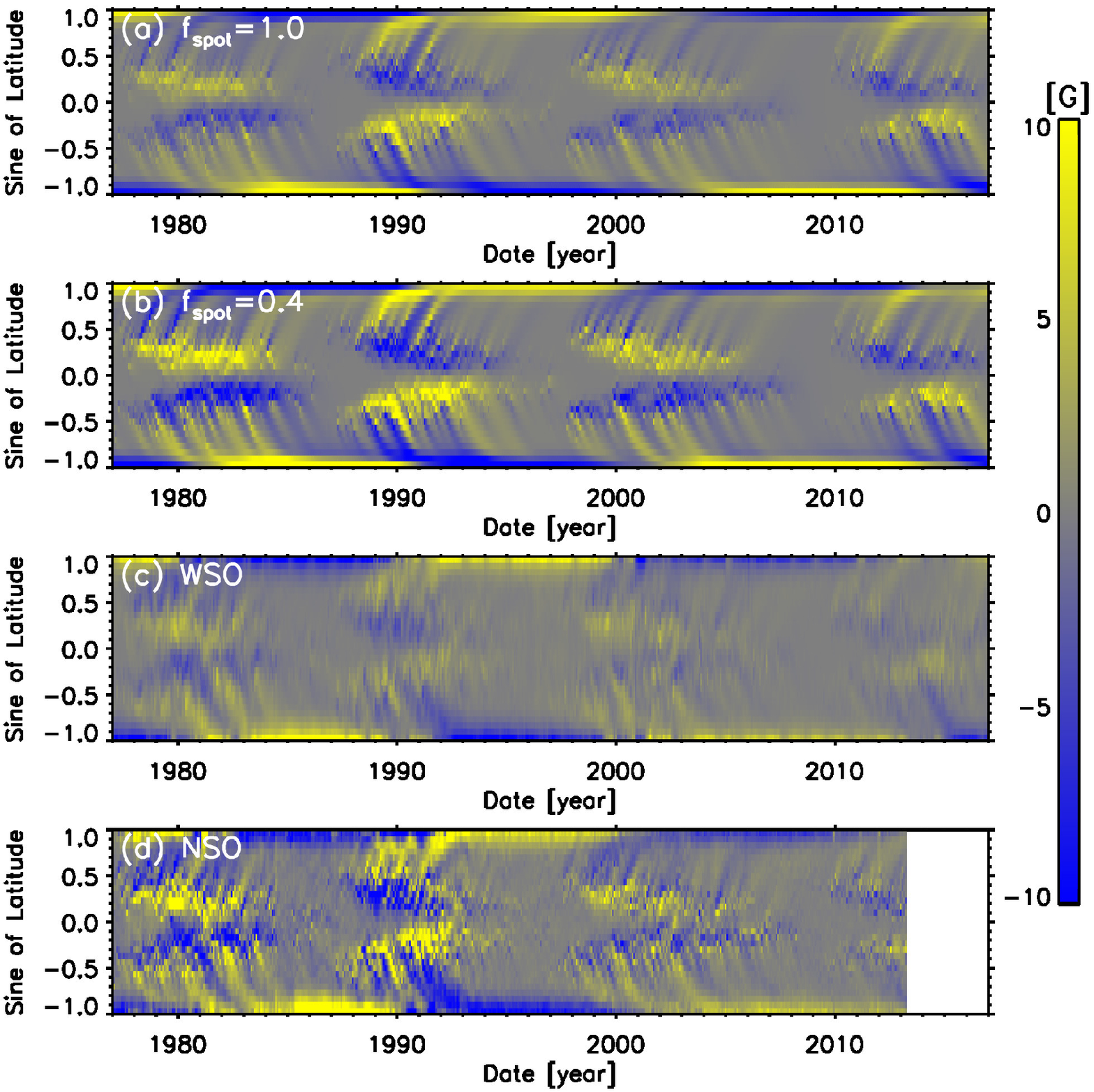}
\caption{
 Magnetic butterfly diagram of the optimal solutions shown in Table \ref{tab:optimal_solution}.
 Illustrated are the cases without asymmetry ($f_\mathrm{spot}=1.0$; panel a), with asymmetry ($f_\mathrm{spot}=0.4$; panel b), the observed magnetogram by WSO (panel c), and the observed magnetogram by NSO (panel d).
 The NSO synoptic magnetogram was calibrated to reproduce the axial dipole strength derived from the WSO data (see the body text in Appendix \ref{appendix:nso} for details).
\label{fig:plmbat_cmp_rgo}
 }
\end{figure}
The area asymmetry of the sunspots provided stronger enhancement to the ``grainy'' structure in the activity belts of the magnetic butterfly diagram (Fig. \ref{fig:plmbat_cmp_rgo}).
As the asymmetry reduces the contribution of the polar magnetic field if the BMR flux is constant as shown in Section \ref{sec:bmr}, greater amount of the magnetic flux is required to realize the observed axial dipole strength.
The increase of the magnetic fluxes in the activity belts is sometimes preferred in the SFT model because the observed butterfly diagram tends to show more ``grainy'' activity belts \citep{2014ApJ...791....5J}.
In addition to the scatter of the tilt angle \citep{2014ApJ...791....5J}, the asymmetry of the leading and following sunspots can be a new candidate that increases the average net unsigned flux at the low latitudes without increasing the net flux at high latitudes.
Both simulations showed stronger magnetic field in the low latitude than in the WSO synoptic magnetogram, which will be caused by the limited spatial resolution of the WSO observation.
For comparison with the observed magnetic butterfly diagram, we used the synoptic magnetogram data of NSO (see Appendix \ref{appendix:nso} for the calibration of the NSO data).
The strong enhancement of the activity belt in Cycles 21 and 22 observed by NSO (panel d) showed better agreement with the SFT simulation of $f_\mathrm{spot}=0.4$ (panel b).
The enhancement of the activity belt observed by NSO (panel d) was weaker in Cycles 23 and 24.
Both simulations showed greater enhancement of the activity belt than the NSO data in Cycle 24.
Considering the difference between the observational instruments, it appears to be difficult to decide which simulation shows better agreement with the observation in these cycles.
Further quantitative analysis should be carried out in future studies.

\begin{figure}[!ht]
\epsscale{0.8}
\plotone{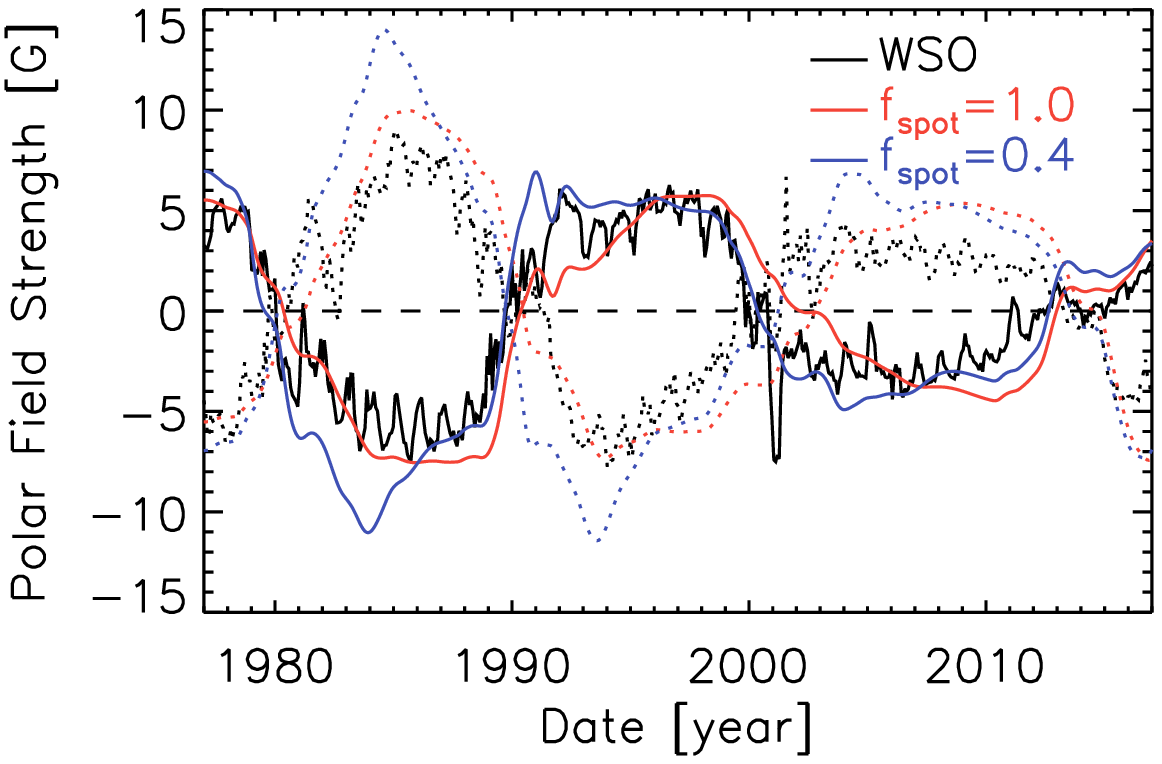}
\caption{
 Polar magnetic field strength averaged over $|\lambda|{\ge}55$ of the optimal solutions shown in Table \ref{tab:optimal_solution}.
 Shown are the cases without asymmetry ($f_\mathrm{spot}=1.0$; red), with asymmetry ($f_\mathrm{spot}=0.4$; blue), and the observation by WSO (black).
 The polar fields in the northern and southern hemispheres are shown by solid and dashed lines, respectively.
\label{fig:plmpf_cmp_rgo}
 }
\end{figure}
As we optimized the solutions by minimizing the difference from the observed axial dipole strength, it was not guaranteed that the optimized solution would also demonstrate good correspondence with the observed polar magnetic field strength.
The solution with the sunspot area asymmetry tended to show the timing of the polar field reversal closer to the observation as seen in Fig. \ref{fig:plmpf_cmp_rgo}.
We should also note that the simulated polar field tended to show greater variation than the observation, especially in Cycle 21.
The possible reasons of this discrepancy will be the optimization procedure focusing only on the axial dipole strength or the inconsistency between the modeled and real sunspot group.

\section{Summary and discussion}
\label{sec:summary}

In this paper, we have studied, for the first time, the effect of the morphological asymmetry between leading and following sunspots on the polar magnetic field by using the surface flux transport model.
We present that the morphological asymmetry of active regions prevents the polar field build-up depending on the latitudinal location and spatial size of BMRs.
As the observational studies on the morphological asymmetry in this context are severely limited as compared to the other asymmetry within a BMR (e.g., Joy's law), we conjecture the BMR parameters from the observed ratio of the sunspot area.
According to the SFT simulations of the last four solar cycles, the inclusion of the morphological asymmetry reduces the root-mean-square difference from the observed axial dipole strength by 30--40 percents.
However, we further observed that the polar magnetic field of the simulation with asymmetric BMRs deviates from the observation near the end of Cycle 21.
As the purpose of this paper is to present the qualitative and quantitative effect of the morphological asymmetry on the polar field formation, we retain this problem for future studies.


The anomalously long cycle minimum between Cycles 23 and 24 and the resulting weak polar field has been a target of broad discussion \citep{2015LRSP...12....5P,2015LRSP...12....4H}.
As the origin of this weak and long duration of the Cycle 23/24 minimum, a wide variety of explanations was investigated such as the abnormal tilt angle near the cycle minimum \citep{2015ApJ...808L..28J}, the active region inflow \citep[e.g.,][]{2002ApJ...570..855H,2004ApJ...603..776Z}, high gradient of the meridional flow at low latitude \citep{2008SoPh..252...19S,2009ApJ...707.1372W,2013SSRv..176..289J}, temporal variation of the meridional flow \citep{2014ApJ...792..142U},
or the global-scale decay of the polar magnetic field \citep[e.g.,][]{2002ApJ...577.1006S,2006A&A...446..307B,2017A&A...604A...8V}.
We observe that the long flat profile of the Cycle 23/24 minimum can be reproduced by including the effect of the morphological asymmetry between the leading and following sunspots (Fig. \ref{fig:plmdp_cmp_rgo}).
The result suggests that the morphological asymmetry may be a new candidate for the origin of the long minimum between Cycle 23 and 24.
A similar flat profile near the cycle minimum is observed in the last few cycles with the axial dipole strength \citep{2017A&A...607L...2I} and polar microwave brightness temperature \citep{2018JASTP.176...26G}.
For older cycles, \cite{2012ApJ...753..146M} provides the polar faculae measurements that do not always show the temporal flatness before the cycle minimum.
Further studies are required to understand the role of the morphological asymmetry in the temporal profile of the polar magnetic field.



We assumed that the sunspot asymmetry parameter $f_\mathrm{spot}$ is constant in time and latitude.
\cite{2014SoPh..289.1403T} reported the time dependence and hemispheric asymmetry of $f_\mathrm{spot}$.
In addition, it is natural to expect that $f_\mathrm{spot}$ exhibits latitudinal dependence from the Colioris force \citep[e.g.,][]{1993ApJ...405..390F,2014ApJ...785...90R} with a random scatter by the convective flow \citep{2011ApJ...741...11W}.
These effects were not considered in our BMR model.
Further observational works should be carried out focusing on the statistical properties of the leading and following asymmetry in sunspots.

In the optimization of the free parameters in Sec. \ref{sec:whole}, we implicitly assumed that all of the cycle dependencies are caused by the cycle-to-cycle difference of the mean tilt angle.
There are various possible sources of the cycle dependencies of the SFT models such as the temporal variation and nonlinearity of the meridional flow \citep[e.g.,][]{2011ApJ...729...80H,2018ApJ...864L...5I}, the randomness of the convective motion \citep[e.g.,][]{2014ApJ...780....5U}, or the long-term variation of the instrument and sunspot data.
The temporal variation of the morphological asymmetry \citep{2014SoPh..289.1403T,2015AdSpR..55..835T} is a possible source of the cycle dependence of the surface flux transport process as well.
There is further scope for the optimization procedures in our study.
More sophisticated optimization procedures \citep[e.g.,][]{2017A&A...607A..76W} should be used in the future models with a larger number of free parameters.

\tred{
In this study, we have attempted to quantify the possible impact of the morphological asymmetry between leading and following sunspots.
However, there still exist large uncertainties, which mainly arise from the lack of enough observational studies on this topic.
We used $\delta_\mathrm{spot}=0.4$ as the typical ratio between the leading and following sunspot areas following \cite{2014SoPh..289.1403T}.
On the other hand, \cite{2014SoPh..289..563M} suggests that the leading sunspot area is typically 25 percents larger than the following area, which corresponds to $\delta_\mathrm{spot}{\sim}0.8$.
Because the sunspot area ratio of 0.8 inferred from \cite{2014SoPh..289..563M} is derived only for the active regions where the number of leading sunspots is equal to the number of following sunspots in each active region.
Because we are interested in the net effect of the morphological asymmetry including the active regions that have only leading spots, we used the results by \cite{2014SoPh..289.1403T}.
More observational studies should be carried out in the future.
The assumption of the Gaussian function is also a source of the uncertainty in this study.
The sunspots do not show complete round shape nor long tail far from the central point as found in the Gaussian function.
The actual active regions are not always bipolar.
The use of more realistic source term based on the observed magnetogram \citep[e.g.,][]{2015SoPh..290.3189Y} will be a possible way to avoid these difficulties including the realistic effect of the morphological asymmetry.
}

\acknowledgements

This work was supported by JSPS KAKENHI Grant Number JP15H05816, JP18H04436, and JP19K14756.
This work was also partially supported by ISEE CICR International Workshop program, and the authors would like to thank all the members of the workshop.
Numerical computations were carried out in the Center for Integrated Data Science, Institute for Space-Earth Environmental Research, Nagoya University through the joint research program.
Wilcox Solar Observatory data used in this study were obtained via the web site \url{http://wso.stanford.edu}.
The Wilcox Solar Observatory is currently supported by NASA.
NSO/Kitt Peak data used here are produced cooperatively by NSF/NOAO, NASA/GSFC, and NOAA/SEL.
Data were acquired by SOLIS instruments operated by NISP/NSO/AURA/NSF. Data storage was supported by the University of Colorado Boulder "PetaLibrary."

\appendix
\section{One-dimensional expression of magnetic patch in the BMR model \label{appendix:1d_source}}

For the simplicity, we consider the longitudinal average of the Gaussian function on a sphere
\begin{equation}
  B(\theta,\phi)=
  \exp\left[-\frac{2\left(1-\cos\beta\right)}{\delta^2}\right],
\end{equation}
where
\begin{equation}
  \cos\beta=
  \cos\theta\cos\theta_0
  +\sin\theta\sin\theta_0\cos\left(\phi-\phi_0\right)
\end{equation}
and $\left(\theta_0,\phi_0\right)$ is the location of the central point of the Gaussian function.
This Gaussian function can be analytically averaged over the longitude
\begin{align}
  B(\theta)&=\frac{1}{2\pi}\int_0^{2\pi}
  \exp\left[-\frac{2\left(1-\cos\beta\right)}{\delta^2}\right]d\phi
  \nonumber\\
&=\exp\left[-\frac{2\left(1-\cos\theta\cos\theta_0\right)}{\delta^2}\right]
  \cdot\frac{1}{2\pi}\int_0^{2\pi}\exp\left[
  \frac{2\sin\theta\sin\theta_0}{\delta^2}
  \cos(\phi-\phi_0)\right]d\phi
  \nonumber\\
&=\exp\left[-\frac{2\left(1-\cos\theta\cos\theta_0\right)}{\delta^2}\right]
  \cdot\mathrm{I}_0\left(\frac{2\sin\theta\sin\theta_0}{\delta^2}\right),
\end{align}
where $I_0$ is the zeroth order modified Bessel function of the first kind.
As each BMR can be expressed as a sum of two Gaussian functions, the above expression can be directly used to obtain the one-dimensional flux emergence term from the two-dimensional BMR model.

The magnetic flux of the Gaussian function is defined by
\begin{equation}
  \Phi=R_\sun^2
  \int_{0}^{2\pi}d\phi\int_{0}^{\pi}\sin\theta{d\theta}
  B(\theta,\phi).
\end{equation}
Without loss of generality, we can set $\theta_0=0$.
The magnetic flux can be derived as
\begin{equation}
  \label{eq:mn_flux}
  \Phi={\pi}R_\sun^2{\delta^2}
  \left[1-\exp\left(-\frac{4}{\delta^2}\right)\right].
\end{equation}
For real sunspots, we can assume that $\delta\ll2$ and we obtain a simpler relation
\begin{equation}
  \Phi{\sim}{\pi}R_\sun^2{\delta^2}.
\end{equation}

\section{Effect of the sunspot diffusion on the leading/following asymmetry}
\label{appendix:asymmetry}

The analytical expression of the diffusing magnetic patch can be written as
\begin{equation}
  B(r,t)=\exp\left[-\left(\frac{r}{\delta(t)}^2\right)\right],
\end{equation}
where
\begin{equation}
  \delta(t)^2=\delta_0^2+4{\eta}{t}
\end{equation}
and $\delta_0$ is the initial size of magnetic patch at $t=0$.

The spatial size of a initially small magnetic patch can be comparable to its latitude $\lambda_\mathrm{c}$ when
\begin{equation}
  t{\ge}\frac{\lambda_\mathrm{c}^2}{4\eta}
  {\sim}171
  \left(\frac{\lambda_\mathrm{c}}{10\ \mathrm{[deg]}}\right)^2
  \left(\frac{250\ \mathrm{[km^2/s]}}{\eta}\right)
  \ \mathrm{[days]}.
\end{equation}
This time scale is comparable to that of the magnetic flux transport by the poleward meridional flow.

The time evolution of the asymmetry can be estimated by this relation.
\begin{equation}
  f_\delta(t)
  =\frac{\delta(t)^{F}}{\delta(t)^{L}}
  =\frac{\left(\delta_0^{F}\right)^2+4{\eta}{t}}
  {\left(\delta_0^{L}\right)^2+4{\eta}{t}}
\end{equation}
This is a monotonic function that approaches $f_\delta=1$.
The result indicates that, without the effect of the flow, flux cancelation, and nonlinear effects, the asymmetry is the maximum when the sunspots were first introduced in the SFT simulation and gradually reduced in time.
The relaxation time scale of the asymmetry can be approximated as
\begin{equation}
  t{\sim}\frac{1}{4\eta}
  \max\left(\delta_0^{L},\delta_0^{F}\right).
\end{equation}
With the magnetic diffusion of 250 km$^2$/s and the spatial size of 4 deg, the time scale becomes approximately 27 days.
The asymmetry of a BMR with greater spatial size continues to be longer and has stronger effect on the polar field formation.

\section{Data calibration between WSO and NSO synoptic maps}
\label{appendix:nso}

In this study, the synoptic magnetogram data observed by National Solar Observatory (NSO) was used for the comparison with the observation in Sec. \ref{sec:whole}.
Here, we describe the calibration of the NSO data such that the axial dipole strength of the NSO and WSO data shows agreement.
The reason why we chose the axial dipole strength observed by the WSO data as a reference of the calibration of the NSO data is because (1) the quality of the WSO data appears to be uniform in time and (2) we used the WSO data for optimizing our SFT simulations in Sec. \ref{sec:whole}.
We do not intend to argue which observation is better or worthwhile.

We combined two synoptic magnetograms observed by the NSO/KPVT (available from 1976 until 2003) and NSO/SOLIS (from 2003).
According to the description file of the KPVT data (\url{ftp://nispdata.nso.edu/kpvt/synoptic/README}), the KPVT synoptic magnetogram map data is the combination of different instruments.
The data was calibrated with the axial dipole strength of WSO data using the least-square method to determine the spatially and temporally uniform correction factor for scaling the synoptic magnetic field data.
For the KPVT data, we discovered a better fitting result by assuming two different correction factors before and after the Carrington rotation number 1863 (around November 1992).
We believe this is reasonable considering the change of observational instruments around 1992.
We used a temporally constant correction factor for SOLIS data.

\begin{figure}[!ht]
\epsscale{0.8}
\plotone{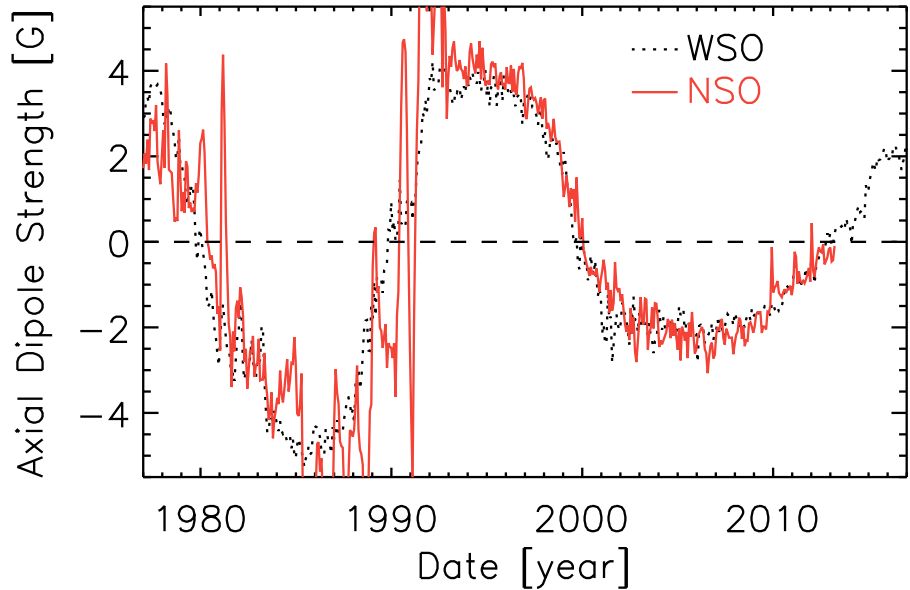}
\caption{
 Axial dipole strength of the WSO (black dotted) and calibrated NSO (red solid) synoptic magnetograms.
\label{fig:plmdp_cmp_obs}
 }
\end{figure}

\begin{figure}[!ht]
\epsscale{0.8}
\plotone{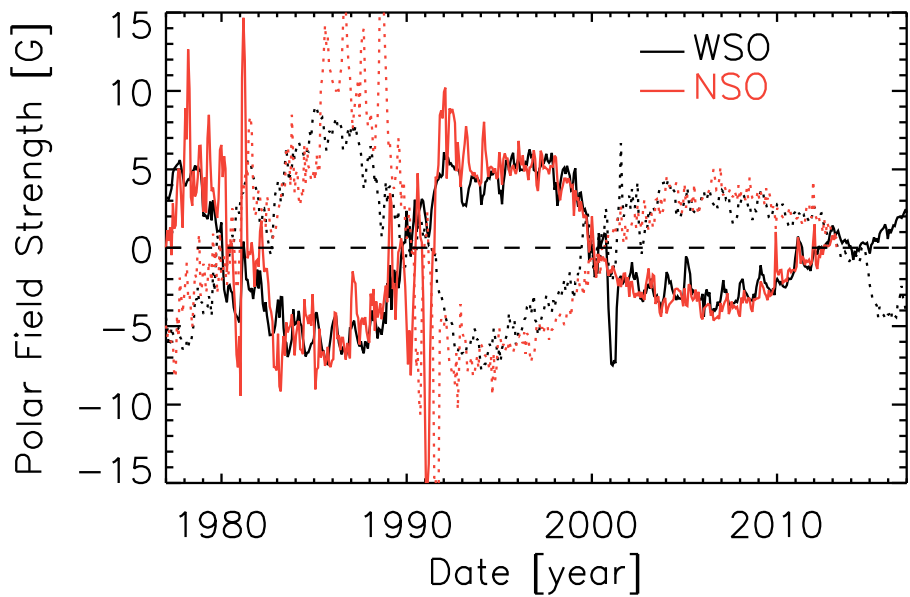}
\caption{
 Polar magnetic field strength averaged over $|\lambda|{\ge}55$ of the WSO (black) and calibrated NSO (red) synoptic maps.
 The polar fields in the northern and southern hemispheres are shown by solid and dashed lines, respectively.
\label{fig:plmpf_cmp_obs}
 }
\end{figure}

Figure \ref{fig:plmdp_cmp_obs} shows the time evolution of the axial dipole strength of the WSO data and calibrated NSO data following the above procedure.
Although the dispersion of the KPVT data exists before 1992, the overall evolution of the axial dipole strength seems to be in good agreement with the WSO data.
Figure \ref{fig:plmpf_cmp_obs} shows the time evolution of the polar magnetic field averaged where the latitude is greater than 55 degrees.
Although the NSO data are not calibrated for the averaged polar magnetic field, the NSO data shows good agreement with the WSO data.
Although better calibration procedures should exist considering the latitudinal or temporal dependence of the correction factors, we believe that the calibrated data has a marginal accuracy for rough comparison on the strength and shape of the magnetic butterfly diagram.
We therefore use this calibrated NSO data as a reference to be compared with the simulated data in Sec. \ref{sec:whole}.

%
%
%
%



\end{document}